\documentclass[reprint,amsmath,amssymb,aps,prl,longbibliography]{revtex4-1}

\usepackage{bm}
\newcommand{\kF}{\ensuremath{k_\mathrm{F}}}
\newcommand{\dd}{\mathrm{d}}
\newcommand{\ii}{\mathrm{i}}
\newcommand{\nd}{{\hphantom{\dagger}}}
\newcommand{\qq}{{\bf{q}}}
\newcommand{\kk}{{\bf{k}}}
\newcommand{\kma}{k_\mathrm{m}^\ast}

\usepackage{graphicx}
\usepackage{xcolor}
\definecolor{col1}{cmyk}{.09,1,0.54,.43} 
\definecolor{col2}{cmyk}{1,.48,.12,.58} 
\definecolor{col3}{cmyk}{1,.22,.02,.18} 

\usepackage[
  colorlinks,
  breaklinks,
  citecolor=col1,
  linkcolor=col2,
  urlcolor=col3,]{hyperref}

\begin{document}

\title{
    Beyond mean-field corrections to the quasiparticle spectrum 
    of superfluid Fermi gases}

\author{Senne Van Loon}
\email{Senne.VanLoon@UAntwerpen.be}
\author{Jacques Tempere}
\author{Hadrien Kurkjian}
\affiliation{TQC, Universiteit Antwerpen, Universiteitsplein 1, B-2610
                Antwerpen, Belgi\"e}

\date{\today}

\begin{abstract}
    We investigate the fermionic quasiparticle branch of superfluid Fermi gases
    in the BCS-BEC crossover and calculate the quasiparticle lifetime and energy
    shift due to its coupling with the collective mode. The only close-to-resonance process
    that low-energy quasiparticles can undergo at zero temperature is the emission of a
    bosonic excitation from the phononic branch. Close to the minimum of
    the branch we find that the quasiparticles remain undamped, allowing us 
    to compute corrections to experimentally
    relevant quantities such as the energy gap, location of the minimum,
    effective mass, and Landau critical velocity. 
\end{abstract}

\maketitle

\textit{Introduction}---
The notion of quasiparticles is an essential tool for the study of interacting
many-body systems. The idea is that the many-body problem can be
considerably simplified by introducing weakly-interacting elementary excitations
above a known ground state. It is, however, generally known that fermionic
quasiparticles in an interacting Fermi system are only well-defined in a small
region around the dispersion minimun, and that elsewhere they acquire a finite
lifetime, even at zero temperature \cite{nozieres}. In a superfluid,
the fermionic branch, which consists of pair-breaking excitations, obtains a
finite energy gap described by BCS theory \cite{BCS1957}, and the system
exhibits a gapless bosonic mode (unlike in superconductors \cite{Anderson1958})
representing the collective motion of fermion pairs \cite{Stringari2006CM}. In
this Letter, we identify the coupling of the fermionic quasiparticle to the
bosonic collective branch as the mechanism responsible for its finite lifetime
away from the dispersion minimum, and compute the corresponding damping rate and energy
shift from first principles. This problem is quite similar to the Bose
polaron (an impurity in a sea of bosons)
\cite{Landau1933,Utesov2018,VanLoon2018}, but a crucial difference is that the
fermionic quasiparticle has a roton-like dispersion
$\epsilon_k=\Delta+\hbar^2(k-k_0)^2/2m^\ast$, rather than an impurity-like
dispersion $\epsilon_k=\hbar^2k^2/2m^\ast$.

The appearance of a finite quasiparticle lifetime away from the energy
minimum is a many-body phenomenon that occurs in many quantum systems such as
normal Fermi liquids \cite{nozieres,Panholzer2012}, superconductors
\cite{Scalapino1976}, or rotonic systems (superfluid Helium \cite{Chernyshev2012}
or dipolar gases \cite{Ferlaino2018}). Using ultracold fermionic atoms
\cite{Bloch2008Review,zwerger2011bcs,Jin2003,Thomas2007,Ketterle2008,Stringari2013,Vale2017},
it can be studied analytically from a first-principle microscopic approach, and
compared to experimental observations. Measurements of the quasiparticle
spectrum using rf-spectroscopy \cite{Ketterle2003RF} and momentum-resolved
rf-spectroscopy \cite{jin2008}, in particular of the quasiparticle gap
\cite{Ketterle2008Gap}, are in fact already available. Moreover, using Feshbach
resonances to tune the interaction strength \cite{Grimm2004}, experiments can
study this effect in the whole range between a weakly-interacting BCS-type
superfluid and a Bose-Einstein Condensate (BEC) of tightly-bound dimers,
including at unitarity where interactions are resonant.

Yet, this problem was somehow overlooked in recent theoretical studies,
concentrating rather on computing the equation-of-state
\cite{Drummond2006,Randeria2008,Giorgini2004},
the order parameter \cite{GMB,pieri2004bcs,pisani2018gap}, or
the bosonic collective mode spectrum \cite{Popov1976,Stringari2006CM,Sinatra2017,Kurkjian2017,Castin2019,Klimin2019}.
Pioneering studies have looked at beyond mean-field corrections to the single-particle
Green's function \cite{Haussmann1993,Kopietz2008Absence,Zwerger2009}, identifying the coupling
to the collective mode as the most important effect \cite{Zwerger2009}, but
could not extract analytically the corrected eigenenergy and damping rate; moreover, such
numerical approaches suffer limitations, predicting in particular a finite
lifetime of the quasiparticles at the dispersion minimum. Here, we aim to fill
this gap by analytically studying the coupling of the fermionic branch with the
bosonic collective mode, modifying its dispersion in the entire BCS-BEC
crossover. As expected, we find the quasiparticles to be well-defined around the
Fermi level, while their lifetime becomes finite away from the energy minimum.
The correction is perturbative in both the BCS and the BEC limit, expressing a
vanishing damping rate in these limits.

\textit{Quasiparticle Hamiltonian}---
We study a gas of neutral fermionic atoms of mass $m$ in two different hyperfine
states, interacting via an attractive short-range potential, fully characterized, at
low energy, by its $s$-wave scattering length $a$. Instead of using the full
microscopic Hamiltonian, we describe the weakly-excited state of the system in
terms of its quasiparticles. Specifically, we use an effective Hamiltonian
derived from first principles that describes the fermionic quasiparticles and
their coupling to the Anderson-Bogoliubov bosonic collective modes
\cite{Kurkjian2017}
\begin{align}
    \hat{H}_\mathrm{qp} =& 
    \sum\limits_{\bf{k},\sigma=\uparrow,\downarrow} 
        \epsilon_{\bf{k}} \hat{\gamma}^\dagger_{\bf{k},\sigma} \hat{\gamma}^\nd_{\bf{k},\sigma}
    +\sum\limits_{\bf{q}} \hbar \omega_{\bf{q}}^\nd 
        \hat{b}^\dagger_{\bf{q}} \hat{b}^\nd_{\bf{q}} \notag \\
    &+\frac{1}{\sqrt{V}} \sum\limits_{\bf{k},\bf{q},\sigma}
        \Big(\mathcal{A}^\nd_{\bf{k}-\bf{q},\bf{q}} \hat{b}_{\bf{q}}^\dagger 
            +\mathcal{A}^\nd_{\bf{k},-\bf{q}} \hat{b}^\nd_{-\bf{q}} \Big)
            \hat{\gamma}^\dagger_{\bf{k}-\bf{q},\sigma} \hat{\gamma}^\nd_{\bf{k},\sigma}.
    \label{eq:Hamiltonian}
\end{align}
Here, the first term describes the BCS quasiparticles with creation
(annihilation) operators $\hat{\gamma}^\dagger_{\bf{k},\sigma}\,(\hat{\gamma}^\nd_{\bf{k},\sigma})$ and energy $\epsilon_{\bf{k}}=
\sqrt{\xi_{\bf{k}}^2+\Delta^2}$, with $\xi_{\bf{k}}=\frac{\hbar^2k^2}{2m}-\mu$
the free-fermion dispersion relation, $\Delta$ the mean-field gap
and $\mu$ the chemical potential. The second term in Eq.~\eqref{eq:Hamiltonian}
represents the free bosonic collective modes, calculated within the RPA or
Gaussian pair fluctuations (GPF) approximation \cite{Anderson1958,
Stringari2006CM, Randeria2008}, with operators $\hat{b}^\dagger_{\bf{q}},\,
\hat{b}^\nd_{\bf{q}}$; their eigenenergy $\hbar \omega_{\bf{q}}$ is the only
positive real root of the gaussian fluctuation matrix $\mathrm{det}
M(\omega_{\bf{q}},{\bf{q}})=0$ \cite{Stringari2006CM}, with
\begin{align}
    M_{\!_{\pm\pm}}^{}\!(\omega,{\bf{q}})\!=& \frac{1}{2V} \kern-.3em\sum\limits_{\bf{k}}
    \bigg[
        \frac{\epsilon_{\bf{k}_+}\!\!+\epsilon_{\bf{k}_-}}
            {\epsilon_{\bf{k}_+} \epsilon_{\bf{k}_-}}
        \frac{\epsilon_{\bf{k}_+}\epsilon_{\bf{k}_-} \!\!
                + \xi_{\bf{k}_+}\xi_{\bf{k}_-} \!\!\pm \Delta^2}
            {\hbar^2 \omega^2-\big( \epsilon_{\bf{k}_+}+\epsilon_{\bf{k}_-} \big)^2}
        + \frac{1}{\epsilon_{\bf{k}}}
    \bigg] \notag\\
    M_{\!_{+-}}^{}\!(\omega,{\bf{q}})\!=& \frac{\hbar \omega}{2V} \sum\limits_{\bf{k}}
        \frac{1}{\epsilon_{\bf{k}_+} \epsilon_{\bf{k}_-}}
        \frac{\epsilon_{\bf{k}_+}\xi_{\bf{k}_-} + \xi_{\bf{k}_+}\epsilon_{\bf{k}_-}}
            {\hbar^2 \omega^2-\big(\epsilon_{\bf{k}_+}+\epsilon_{\bf{k}_-}\big)^2}
    \label{GPFmatrix}
\end{align}
and ${\bf{k}}_\pm = {\bf{k}}\pm{\bf{q}}/2$. Finally, the second line of
Eq.~\eqref{eq:Hamiltonian} describes the three-body coupling between the
fermionic quasiparticles and the collective modes, with coupling
amplitude
\begin{equation}
    \mathcal{A}_{\bf{k},\bf{q}} = 
        \frac{w^+_{\bf{k}\bf{q}} \sqrt{M_{++}(\omega_{\bf{q}},\bf{q})} 
            + w^-_{\bf{k}\bf{q}} \sqrt{M_{--}(\omega_{\bf{q}},\bf{q})}}
        {\sqrt{-\frac{2}{\hbar} \frac{\partial}{\partial \omega} 
            \mathrm{det} M(\omega,\bf{q}) \big|_{\omega=\omega_{\bf{q}}}}},
    \label{eq:CoupAmp}
\end{equation}
with the weights $w^\pm_{\bf{k}\bf{q}} =U_{{\bf{k}}+{\bf{q}}}V_{\bf{k}}\pm
U_{\bf{k}}V_{{\bf{k}}+{\bf{q}}}$, and
$U_{\bf{k}}=\sqrt{(1+{\xi_{\bf{k}}}/{\epsilon_{\bf{k}}})/2},\,
V_{\bf{k}}=\sqrt{(1-{\xi_{\bf{k}}}/{\epsilon_{\bf{k}}})/2}$. The effective
Hamiltonian~\eqref{eq:Hamiltonian} describes accurately the emission or absorption of one
Anderson-Bogoliubov boson. In principle, it also contains higher order decay
processes such as the successive emission of two bosons, but to include
those higher order terms consistently one should consider a three-boson
coupling Hamiltonian~\cite{err}, yet unknown in literature on Fermi condensates.
We also omit highly off-resonant processes such as four-fermion interactions (at
$T=0$: $\gamma^\dagger\gamma^\dagger\gamma^\dagger \gamma$ or
$\gamma^\dagger\gamma^\dagger\gamma^\dagger\gamma^\dagger$) and simultaneous
creation of two fermions and a boson ($\gamma^\dagger\gamma^\dagger b^\dagger$),
that lead to a nonzero contribution to the damping rate only when the
quasiparticle energy is at least $3\Delta$. The process described by
$\hat{H}_\mathrm{qp}$ is the most relevant damping process at zero
temperature and low energy.

\textit{Energy corrections}---
To study the fermionic branch we compute its Green's function \cite{CohenChapIIIC}, 
which contains all the information on the quasiparticle. At zero temperature, the collective
modes are unoccupied, forbidding the absorption of a boson by the fermionic
quasiparticle. In this way, the only second order diagram contributing to the
Green's function is the self-energy where the fermion emits and reabsorbs a boson:
\begin{equation}
    G_{\bf{k}}(z) = 
    \Bigg( z-\epsilon_{\bf{k}} - \frac{1}{V} \sum\limits_{\bf{q}} 
            \frac{|\mathcal{A}_{\bf{k}-\bf{q},\bf{q}}|^2}
                {z-\epsilon_{\bf{k}-\bf{q}}-\hbar \omega_{\bf{q}}} 
    \Bigg)^{-1}.
    \label{eq:GreensF}
\end{equation}
As explained at the end of this Letter, one obtains the same quasiparticle Green's
function within the formalism of Refs.~\cite{Haussmann1993,Zwerger2009} provided
one omits all far-off-shell processes. The poles of the Green's function
$G_{\bf{k}}^{-1}(z_{\bf{k}})=0$ are the eigenenergies of fermionic
quasiparticles dressed  by their interactions with the boson bath. When the
coupling amplitude $\mathcal{A}$ is small~\footnote{At strong-coupling there is
no small parameter guaranteeing the smallness of the coupling amplitude.
However, the correction to the eigenenergy never exceeds 15\%, see e.g.
Fig~\ref{fig:FitParam}(c), and the self-consistent result remains close to the
perturbative one. This suggests that higher order decay processes where two or
more bosons are emitted would lead to even smaller corrections.}, one can
replace $z$ by $\epsilon_{\bf{k}} + \ii 0^+$ in the last term between brackets
of Eq.~\eqref{eq:GreensF}, to obtain the energy correction
$z_{\bf{k}}^{(2)}=E^{(2)}_{\bf{k}}-\ii\hbar \Gamma_{\bf{k}}/2$ to second order
perturbation theory:
\begin{align}
    E^{(2)}_{\bf{k}} &= \epsilon_{\bf{k}} 
        + \frac{1}{V} \,\mathcal{P}\!\sum\limits_{\bf{q}} 
        \frac{|\mathcal{A}_{\bf{k}-\bf{q},\bf{q}}|^2}
        {\epsilon_{\bf{k}}-\epsilon_{\bf{k}-\bf{q}}-\hbar\omega_{\bf{q}}},
        \label{eq:PertEn}\\
    \hbar \Gamma_{\bf{k}} &= \frac{2\pi}{V} \sum\limits_{\bf{q}} 
        |\mathcal{A}_{\bf{k}-\bf{q},\bf{q}}|^2 \,
        \delta\big(\epsilon_{\bf{k}}
            -\epsilon_{\bf{k}-\bf{q}}-\hbar\omega_{\bf{q}}\big).
        \label{eq:Damping}
\end{align}
The resonance condition
$\epsilon_{\bf{k}}-\epsilon_{\bf{k}-\bf{q}}-\hbar\omega_{\bf{q}}=0$ in
\eqref{eq:Damping} is satisfied provided $\epsilon_{\bf{k}}$ is inside the boson
emission continuum $\{\epsilon_{\bf{k}-\bf{q}} + \hbar\omega_{\bf{q}},\bf{q}\}$,
that is, strictly superior to the threshold energy
$\epsilon_\mathrm{th}=\min_{\bf{q}} [\epsilon_{\bf{k}-\bf{q}} +
\hbar\omega_{\bf{q}}]$, which is also the lower edge of the branch cut of
$G_{\bf{k}}(z)$.

Close to the minimum $k_0$ of the unperturbed fermionic branch, the group
velocity of the quasiparticle is smaller than the sound velocity $c$ of the
collective mode $|\partial\epsilon_{{k}}/\partial k|<\hbar c$. For these values
of $k$, the minimum $\min_{u} [\epsilon_{\bf{k}-\bf{q}} + \hbar\omega_{\bf{q}}]$
over the scattering angle $u=\kk\cdot\qq/kq$ is a strictly increasing function
of $q$, starting from its lowest value $\epsilon_{\bf{k}}$ in $q=0$, such that
$\epsilon_\mathrm{th}=\epsilon_\kk$ and the decay by emission of collective
modes is energetically forbidden \cite{Zwerger2009,Kurkjian2017}. The
perturbative damping rate is zero ($\Gamma_{\bf{k}}=0$) and we find
correspondingly a real pole of $G_\kk$, indicating that the quasiparticles,
despite their renormalisation by the bosonic bath, remain well-defined close to
the minimum of their dispersion.

When the group velocity $|\partial\epsilon_{{k}}/\partial k|$ becomes larger
than $\hbar c$ (which can happen in both the increasing [$k>k_0$] and decreasing
parts [$k<k_0$] of the BCS branch), the unperturbed energy $\epsilon_{\bf{k}}$
becomes greater than $\epsilon_\mathrm{th}$ and the resonance condition of
Eq.~\eqref{eq:Damping} can be satisfied. Although the perturbative damping rate
$\Gamma_\kk$ becomes nonzero, the self-consistent solution $z_{\bf{k}}$ below
$\epsilon_\mathrm{th}$ remains real for larger values of $|k-k_0|$. Eventually,
$z_{\bf{k}}$ also enters the continuum and becomes imaginary, resulting in a
broadened peak in the spectral function $\varepsilon\mapsto
\mbox{Im}[G_\kk(\varepsilon+\ii0^+)]$.

\begin{figure}
    \centering
    \includegraphics{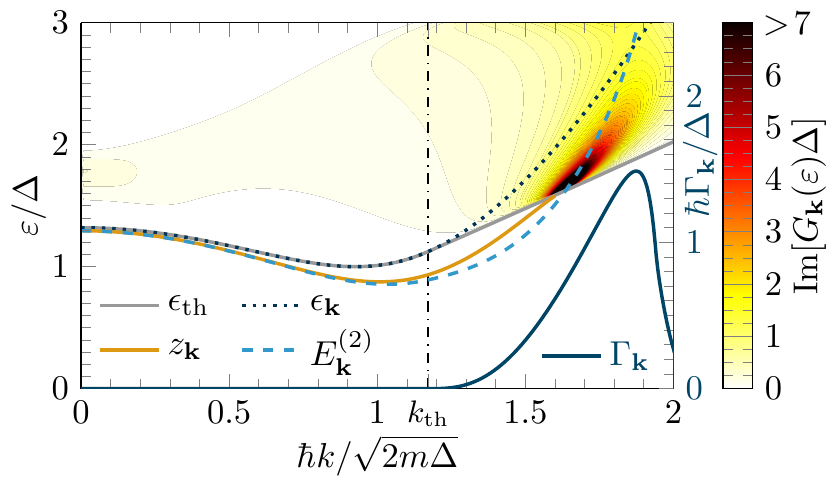}
    \caption{
        (Color online) The imaginary part of the quasiparticle Green's function
        \eqref{eq:GreensF} is shown as a function of the wavenumber $k$
        and energy $\varepsilon$ in units of $\Delta$. It is nonzero only above
        the threshold energy $\epsilon_\mathrm{th}$ (solid gray line).
        Black dotted line: mean field energy $\epsilon_{\bf{k}}$, blue dashed
        line: perturbative eigenfrequency  $E^{(2)}_{\bf{k}}$, 
        blue solid line: perturbative damping rate $\Gamma_\kk$. 
        The latter is nonzero only for $k>k_{\rm th}\simeq 1.17
        \sqrt{2m\Delta}/\hbar$ (vertical dashed dotted line)
        with $|\partial\epsilon_{\bf{k}}/\partial k|_{k=k_{\rm th}}=\hbar
        c$. Below the threshold energy, a self-consistent energy $z_{\bf{k}}$
        can be found until $ k \simeq 1.62 \sqrt{2m\Delta}/\hbar>k_{\rm th}$
        (orange solid line). It remains relatively close to the perturbative
        energy.
    }
    \label{fig:UniSD}
\end{figure}

In Fig.~\ref{fig:UniSD}, we illustrate this by plotting  $\varepsilon\mapsto
\mbox{Im} G_\kk(\varepsilon+\ii0^+)$ at unitarity ($1/\kF a = 0$, where $\kF$ is
the Fermi wavenumber) as a function of $\varepsilon$ and $k$. We superimpose the
eigenenergy $z_{\bf{k}}$ obtained by self-consistently solving for the real pole
of the Green's function below $\epsilon_\mathrm{th}$, and the perturbative
result $E^{(2)}_{\bf{k}}$ of Eq.~(\ref{eq:PertEn}), which remains fairly close
to $z_\kk$ everywhere. Once the self-consistent solution hits the continuum, the
exact resonance of $G_\kk$ turns into a broadened peak at energies
$\varepsilon>\epsilon_{\rm th}$. The perturbative damping $\Gamma_{\bf{k}}$
becomes nonzero at $k=k_{\rm th}$ with $|\partial\epsilon_{\bf{k}}/\partial
k|_{k=k_{\rm th}}=\hbar c$. It is then highly peaked when the energy
$\epsilon_\kk$ is around $3\Delta$. This could suggest that the $1\to3$ fermionic processes
that we excluded from our Hamiltonian $\hat{H}_\mathrm{qp}$ in
Eq.~\eqref{eq:Hamiltonian} become important above $3\Delta$. At higher wavenumbers,
$E^{(2)}_{\bf{k}}$ approaches the mean field result (as does the maximum of
$\mbox{Im}[G_\kk(\varepsilon+\ii0^+)]$). This is not a surprise since the
coupling $\mathcal{A}$ is comparatively small in the limit $k\to\infty$.

\textit{BCS and BEC limits}---
The perturbative treatment Eqs.~\eqref{eq:PertEn}--\eqref{eq:Damping} (already
close to the self-consistent solution at unitarity) gives asymptotically the
exact solution of $G_\kk^{-1}=0$ in the BCS and BEC limits ($\Delta/|\mu|
\rightarrow 0$). In the BCS limit, the bosonic wavenumbers should be expressed
in units of $\Delta/\hbar c$, such that the dispersion $\hbar
\omega_{\bf{q}}/\Delta$ becomes a universal function of $\tilde{\bf{q}}=\hbar c
{\bf{q}}/\Delta$ \cite{kurkjian2016concavity}. The energy correction
$|z_{\bf{k}}-\epsilon_{\bf{k}}|$ is of order $\Delta^2/\mu^2$:
\begin{equation}
    \frac{z_{\bf{k}}^{(2)}\!-\epsilon_{\bf{k}}}{\Delta} = 
        \frac{\Delta^2}{\mu^2} \frac{1}{V}\!\sum\limits_{\tilde{\bf{q}}} 
            \frac{(w^-_{{\bf{k}}\bf{q}})^2}
                {K(\tilde{\bf{q}})}
            \frac{\Delta}
                {\epsilon_{\bf{k}}-\!\epsilon_{{\bf{k}}-\bf{q}}
                    -\!\hbar \omega_{\bf{q}}+\!\ii0^+},
    \label{eq:BCSlimit}
\end{equation}
with $K(\tilde{\bf{q}})=-32\hbar\Delta/(3\sqrt{3}m\kF) \partial
M_{\!_{++}}/\partial\omega\vert_{\omega=\omega_{\bf{q}}}$ a universal function
of $\tilde{\bf{q}}$. 
In the BEC limit, the energy
should be scaled to the chemical potential $|\mu|$ and analytic results are
available for the collective mode dispersion \cite{Stringari2006CM} at arbitrary
momentum $q$. Consequently, the energy $E^{(2)}_{\bf{k}}$ and damping rate
$\Gamma_{\bf{k}}$ of Eqs.~(\ref{eq:PertEn}-\ref{eq:Damping}) can be computed
analytically, resulting in
\begin{equation} 
    \frac{z_{\bf{k}}^{(2)}-\epsilon_{\bf{k}}}{|\mu|} =
        \frac{\Delta^2}{|\mu|^2}
        \frac{2 \tilde{k}^2-6-8 \ii \tilde{k}}{\tilde{k}^4+10\tilde{k}^2 +9},
    \label{eq:BEClimit}
\end{equation}
with $\tilde{k}=\hbar k/\sqrt{2m|\mu|}$. The quasiparticle lifetime thus
diverges like ${\mu^2}/{\Delta^2}$.

\begin{figure}
    \centering
    \includegraphics{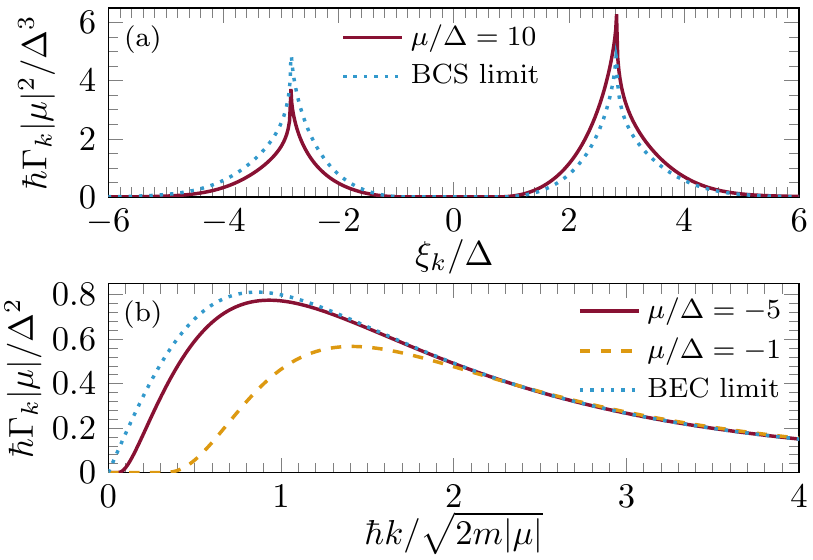}
    \caption{
        (Color online) Perturbative damping rate $\Gamma_{\bf{k}}$ \eqref{eq:Damping}
        for different interaction regimes. (a)
        The damping rate as a function of $\xi_k=\hbar^2k^2/2m-\mu$ at
        $\mu/\Delta=10$ (full red line) is rescaled to compare with the
        universal behavior in the BCS limit (dotted blue line), where the
        damping rate vanishes as $\Delta^2/\mu^2$, according to
        Eq.~\eqref{eq:BCSlimit}. (b) Damping rate in the BEC
        regime as a function of $\hbar k/\sqrt{2m|\mu|}$ for $\mu/\Delta=-5$
        (full red line), and $\mu/\Delta=-1$ (dashed orange line) and $\mu/\Delta=-\infty$
        (blue line, see Eq.~\eqref{eq:BEClimit}).
    }
    \label{fig:Damping}
\end{figure}

In Fig.~\ref{fig:Damping} we examine the perturbative damping rate
$\Gamma_{\bf{k}}$ of Eq.~\eqref{eq:Damping} in the BCS and BEC limits. In the
BCS limit, the undamped region lies around $k_0\approx\kF$ and its width in
units of $\kF$ tends to zero like $\Delta/\varepsilon_{\rm F}$. Outside this
region, a highly peaked behavior can be observed, which, as in the unitary case,
occurs when the energy is close to $3\Delta$. In the BEC limit, the undamped
region lies around $k_0=0$, while the threshold wavenumber vanishes as $\hbar
k_\mathrm{th}/\sqrt{2m|\mu|}=\Delta/4|\mu|$. As a function of $k$, the damping
rate smoothens, and exhibits a $1/k^3$ tail at high $k$.

\begin{figure}
    \centering
    \includegraphics{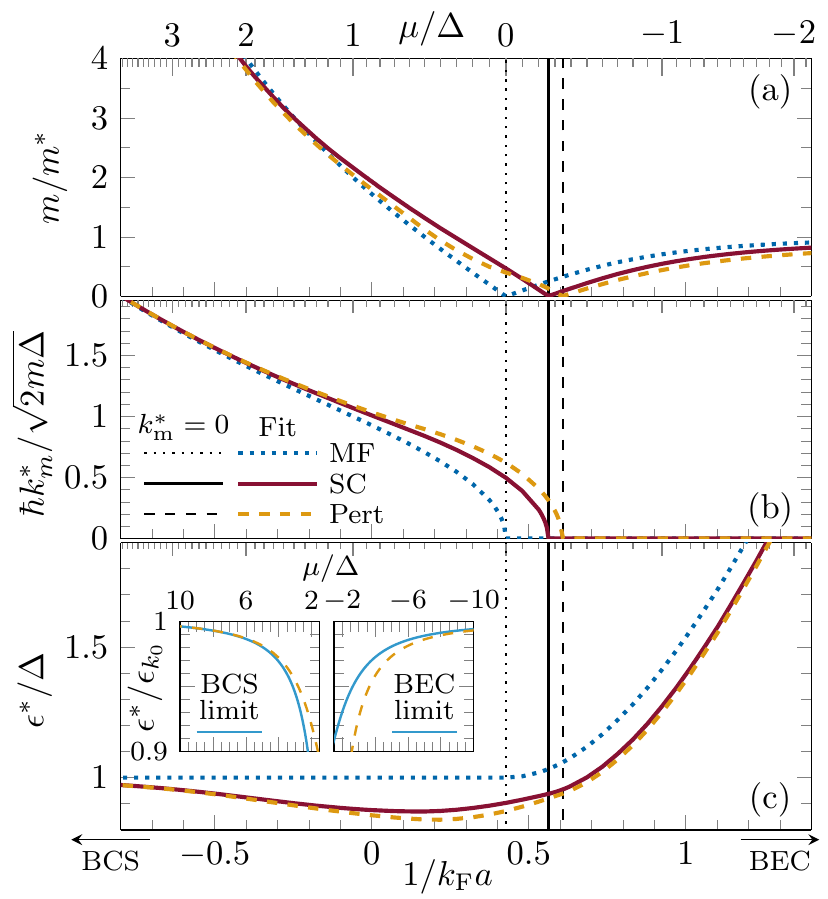}
    \caption{
        (Color online) Minimum of the corrected fermionic energy branch
        fitted to a quadratic dispersion. (a) Dimensionless inverse
        effective mass $m/m^\ast$, (b) location of the minimum $\hbar
        k_m^\ast/\sqrt{2m\Delta}$, (c) energy gap $\epsilon^\ast/\Delta$,
        shown in function of the inverse scattering length $1/\kF a$
        (bottom $x$-axis) and $\mu/\Delta$ (top
        $x$-axis). Full curves: fitting parameters to the
        self-consistent (SC) energy solution $z_{\bf{k}}$. Dashed
        lines: perturbative (Pert) results. Dotted lines:
        mean-field (MF) results. The vertical black lines
        specify the critical values of the interaction after which the minimum
        of the energy is located at $k=0$, in their respective line styles.
        The BCS and BEC asymptotes are drawn for the energy gap in
        inset.
    }
    \label{fig:FitParam}
\end{figure}

\textit{Quadratic dispersion near the minimum}---
To analyze the characteristics of the corrected fermionic energy branch, we fit
a quadratic dispersion
\begin{equation}
    \epsilon_k^\mathrm{fit} = \epsilon^\ast 
        + \frac{\hbar^2 (k-\kma)^2}{2 m^\ast}
    \label{eq:EnFit}
\end{equation}
to the minimum of the energy. This allows us to extract the most interesting
features of the energy correction, in the region where the description in terms
of quasiparticles is certainly accurate. Concretely, these parameters represent
the effective energy gap $\epsilon^\ast$, the location of the minimum of the
branch $\kma$ and the effective mass $m^\ast$. In Fig.~\ref{fig:FitParam} we
present these fitting parameters in the BCS-BEC crossover for both the
self-consistent energy solution and the perturbative energy correction,
comparing with the mean-field version of Eq.~\eqref{eq:EnFit} obtained by
expanding the BCS energy $\epsilon_{\bf{k}}$ around its minimum. All parameters
tend to the mean-field result in the BCS and BEC limits, confirming that the
energy correction is perturbative for $\Delta/|\mu| \rightarrow 0$.
Additionally, the differences between the self-consistent and perturbative
results are never substantial, although perturbation theory somewhat
overestimates the correction. The interaction with the bosonic collective mode
lowers the fermionic energy gap $\epsilon^\ast$, which is to be expected as it
is generally known that the mean-field theory exaggerates the gap. At unitarity,
we find $\epsilon^\ast \simeq 0.88\Delta \simeq 0.41 \epsilon_\mathrm{F}$ (using
the GPF equation of state \cite{Randeria2008}), close to the experimental result
$\Delta = 0.44 \epsilon_{F}$ \cite{Ketterle2008Gap}, and $\kma \simeq
1.01\sqrt{2m\Delta}/\hbar \simeq 0.69 \,\kF$. Furthermore, the
location $\kma$ of the energy minimum reaches $0$ at some critical value
$\mu/\Delta \simeq -0.26$, depicted by a vertical line on
Fig.~\ref{fig:FitParam}, corresponding to $1/\kF a \simeq 0.56$. The fact that
this happens while the chemical potential is already negative has been
theoretically predicted before \cite{Zwerger2009}.

\textit{Critical velocity}---
Another interesting characterization of the fermionic branch is its Landau
critical velocity, which determines the maximal velocity for frictionless flow
at $T=0$ in a superfluid. As there are two branches of elementary excitations in
a superfluid Fermi gas, the critical velocity will be given by the smallest of
the two velocities $v_\mathrm{c} = \min[v_\mathrm{f},c]$, with $v_\mathrm{f} =
\min_{{\bf{k}}_1,{\bf{k}}_2} (\varepsilon_{{\bf{k}}_1} +
\varepsilon_{{\bf{k}}_2})/\hbar|{\bf{k}}_1+{\bf{k}}_2|$ the fermionic critical
velocity \cite{castin2015vitesse}. In the BCS limit
the critical velocity is reached for $k$ close to $\kF$ and we can use the
effective quadratic dispersion \eqref{eq:EnFit} near the minimum to compute it.
This yields $m^*v_{\rm f}^{\rm quad}=  \sqrt{2m^*\epsilon^*+{(\hbar
\kma)^2}}-{\hbar \kma}$, from which we can extract the first deviation to the
mean-field velocity $\frac{v_{\rm f}}{v_\mathrm{f}^{\rm MF}} =
\frac{\epsilon^*}{\Delta}+O(\frac{\Delta}{\mu})^3 \simeq 1 - 0.5
(\frac{\Delta}{\mu})^2 +O(\frac{\Delta}{\mu})^3$.

\textit{Quasiparticle Green's function}---
Finally, we show that the quasiparticle Green's function
Eq.~\eqref{eq:GreensF} can be rederived
in a more general microscopic formalism \cite{Haussmann1993,Zwerger2009,Strinati2019} whose
starting point are the particle, hole and anomalous Green's functions. The
lowest order self-energy correcting the mean-field particle-hole Green's
function
\begin{widetext}
\begin{equation}
    \mathcal{G}_0(z,{\bf{k}})=-\frac{1}{z+\epsilon_{\bf{k}}}
    \begin{pmatrix}
        V_{\bf{k}}^2 & U_{\bf{k}}V_{\bf{k}} \\
        U_{\bf{k}}V_{\bf{k}} & U_{\bf{k}}^2
    \end{pmatrix} + \frac{1}{z-\epsilon_{\bf{k}}}
    \begin{pmatrix}
        -U_{\bf{k}}^2 & U_{\bf{k}}V_{\bf{k}} \\
        U_{\bf{k}}V_{\bf{k}} & -V_{\bf{k}}^2
    \end{pmatrix}
    \label{G0}
\end{equation}
was derived in Ref.~\cite{Haussmann1993}: $\mathcal{G}^{-1} = \mathcal{G}_0^{-1} - \Sigma$, with
\begin{equation}
    \Sigma_{\alpha\alpha^\prime}(z,{\bf{k}}) = 
    \frac{1}{V}\sum\limits_{\bf{q}} \int\limits_{-\ii \infty}^{+\ii\infty} 
    \frac{\dd{z_q}}{2\pi\ii}
        \mathcal{G}_{0,\alpha^\prime\alpha}(z_q-z,{\bf{q}}-{\bf{k}})
        \tilde{M}^{-1}_{\alpha\alpha^\prime}(z_q,{\bf{q}}).
    \label{eqApp:SelfEn}
\end{equation}
Here, $\alpha,\alpha'=1\,\text{or}\,2$ and $\tilde{M}$ is a unitary transform of $M$
defined in Eq.~\eqref{GPFmatrix}
\begin{equation}
    \tilde{M} = -\frac{1}{2}
    \begin{pmatrix}
        M_{++} + M_{--} + 2 M_{+-} & M_{++} - M_{--} \\
        M_{++} - M_{--} & M_{++} + M_{--} - 2 M_{+-}
    \end{pmatrix}.
\end{equation}
This self-energy contains corrections from four-fermion processes (the
lowest order processes allowed for fermions), as can be seen by expanding
$\tilde{M}^{-1}$ in powers of the coupling constant $g_0=-\left(\sum_{\bf{k}}
1/2\epsilon_k\right)^{-1}$. To correctly describe the bosonic spectrum, two of
the intermediate fermions are propagated using the ladder-resummed pair
propagator $\tilde{M}^{-1}$, see Fig.~7 in Ref.~\cite{Zwerger2009}. To obtain
analytic results, and avoid spurious effects (such as the appearance of a
nonzero imaginary part of $\Delta$ at $T=0$), we have dropped the
self-consistent treatment of Ref.~\cite{Zwerger2009} and used in
Eq.~\eqref{eqApp:SelfEn} BCS/RPA expressions for
$\mathcal{G}_0,\,\tilde{M}^{-1}$. At $T=0$, $\tilde{M}^{-1}$ has two real poles
$\pm\hbar\omega_{\bf{q}}$, and two gapped branch cuts. The Matsubara frequency
integral in Eq.~\eqref{eqApp:SelfEn} can then be performed by closing the
contour on the real axis, (the most straightforward way is to avoid the poles of
$\mathcal{G}_0$). Neglecting the far-off-shell contributions of the branch cuts
of $\tilde{M}^{-1}$, and performing the orthogonal transform to the
quasiparticle basis $\tilde{\mathcal{G}}^{-1} = P \mathcal{G}^{-1} P^\dagger$
with
\begin{equation}
    P = \begin{pmatrix}
        U_{\bf{k}} & -V_{\bf{k}} \\
        V_{\bf{k}} &  U_{\bf{k}}
    \end{pmatrix}
\end{equation}
we obtain the corrected Green's function
\begin{equation}
    \tilde{\mathcal{G}}^{-1}(z,\kk) = 
    \begin{pmatrix}
        -z+\epsilon_{\bf{k}} + \frac{1}{V}\sum\limits_{\bf{q}} \left[
            \frac{\mathcal{A}^2_{{\bf{k}}-{\bf{q}},{\bf{q}}}}
                {z-\epsilon_{{\bf{k}}-{\bf{q}}}-\hbar\omega_{\bf{q}}}
            -\frac{\mathcal{B}^2_{{\bf{k}},{\bf{q}}}}
            {-z-\epsilon_{{\bf{k}}+{\bf{q}}}-\hbar\omega_{\bf{q}}}
        \right] & \frac{1}{V}\sum\limits_{\bf{q}} \left[
            \frac{\mathcal{A}_{{\bf{k}}-{\bf{q}},{\bf{q}}}\mathcal{B}_{{\bf{k}}-{\bf{q}},{\bf{q}}}}
                {z-\epsilon_{{\bf{k}}-{\bf{q}}}-\hbar\omega_{\bf{q}}}
            +\frac{\mathcal{A}_{{\bf{k}},{\bf{q}}}\mathcal{B}_{{\bf{k}},{\bf{q}}}}
            {-z-\epsilon_{{\bf{k}}+{\bf{q}}}-\hbar\omega_{\bf{q}}}
        \right] \\ \frac{1}{V}\sum\limits_{\bf{q}} \left[
            \frac{\mathcal{A}_{{\bf{k}}-{\bf{q}},{\bf{q}}}\mathcal{B}_{{\bf{k}}-{\bf{q}},{\bf{q}}}}
                {z-\epsilon_{{\bf{k}}-{\bf{q}}}-\hbar\omega_{\bf{q}}}
            +\frac{\mathcal{A}_{{\bf{k}},{\bf{q}}} \mathcal{B}_{{\bf{k}},{\bf{q}}}}
            {-z-\epsilon_{{\bf{k}}+{\bf{q}}}-\hbar\omega_{\bf{q}}}
        \right] & -z-\epsilon_{\bf{k}} + \frac{1}{V}\sum\limits_{\bf{q}} \left[
            \frac{\mathcal{B}^2_{{\bf{k}}-{\bf{q}},{\bf{q}}}}
                {z-\epsilon_{{\bf{k}}-{\bf{q}}}-\hbar\omega_{\bf{q}}}
            -\frac{\mathcal{A}_{{\bf{k}},{\bf{q}}}^2 }
            {-z-\epsilon_{{\bf{k}}+{\bf{q}}}-\hbar\omega_{\bf{q}}}
        \right]
    \end{pmatrix}.
            \label{Gtilde}
\end{equation}
\end{widetext}
Introducing the weights $W^\pm_{\bf{k}\bf{q}}
=U_{{\bf{k}}+{\bf{q}}}U_{\bf{k}}\pm V_{\bf{k}}V_{{\bf{k}}+{\bf{q}}}$, the
amplitude $\mathcal{B}$ is given by 
\begin{align}
    \mathcal{B}_{{\bf{k}},{\bf{q}}} &=
    \frac{W^+_{\bf{k}\bf{q}} \sqrt{M_{--}(\omega_{\bf{q}},\bf{q})} 
        - W^-_{\bf{k}\bf{q}} \sqrt{M_{++}(\omega_{\bf{q}},\bf{q})}}
    {\sqrt{-\frac{2}{\hbar} \frac{\partial}{\partial \omega} 
        \mathrm{det} M(\omega,\bf{q}) \big|_{\omega=\omega_{\bf{q}}}}}, 
\end{align}
while $\mathcal{A}_{{\bf{k}},{\bf{q}}}$ coincides with Eq.~\eqref{eq:CoupAmp}.
The quasiparticle energy is given by the poles of $ \tilde{\mathcal{G}}(z)$,
hence the roots of $\mathrm{det}\,  \tilde{\mathcal{G}}^{-1}(z)$. Note the
property $\mathrm{det}\,  \tilde{\mathcal{G}}^{-1}(-z)=\mathrm{det}\,
\tilde{\mathcal{G}}^{-1}(z)$, ensuring that the quasiparticle and quasihole energies are
simply opposite. To be consistent with the
omission of the branch cut contributions, we drop in Eq.~\eqref{Gtilde} the
terms corresponding to the far-off-shell processes
$\hat\gamma_{-\kk-\qq,\sigma}^\dagger \hat\gamma_{\kk,\sigma}^\dagger \hat b_{\qq} ^\dagger$,
and restrict to first order in
$\Sigma$ {(which amounts to setting $\mathcal{B}=0$)}. Then $\mathrm{det}\,  \tilde{\mathcal{G}}^{-1}(z)$ is
simply the product of the inverse quasiparticle
$\tilde{\mathcal{G}}_{11}(z)=-G_{\bf{k}}(z)+\mathcal{O}(||\Sigma||^2)$ and
quasihole $\tilde{\mathcal{G}}_{22}(z)=G_{\bf{k}}(-z)+\mathcal{O}(||\Sigma||^2)$
Green's functions
\begin{equation}
    \mathrm{det}\mathcal{G}^{-1} = 
    - G^{-1}_{\bf{k}}(z) G^{-1}_{\bf{k}}(-z) + \mathcal{O}\left(||\Sigma||^2\right).
\end{equation}
This finally explains the use of Eq.~\eqref{eq:GreensF} to study the perturbed
quasiparticle spectrum.

\textit{Conclusion}---
We have corrected the fermionic quasiparticle branch by including its
interaction with the bosonic collective mode. At low energy, this is the only
relevant decay channel, thus giving the sole contribution to the damping of the
single-particle excitations. We have computed this damping rate in the entire
BCS-BEC crossover, and found real poles of the corrected Green's function close
to the minimum of the branch, indicating well-defined quasiparticles.

The boson-emission process we have studied occurs in systems where rotons are
present, such as superfluid Helium \cite{Chernyshev2012} and dipolar Bose gases
\cite{Ferlaino2018}, and should be responsible at $T=0$ for both a shift of the
roton gap and a finite lifetime away from the roton minimum. It occurs also in
normal Fermi liquids \cite{nozieres,Panholzer2012} when the quasiparticles (in
this case gapless) are coupled to the zero sound branch. In superconductors, the
Bogoliubov-Anderson branch is gapped due to Coulomb interactions, but the
quasiparticle can emit crystal phonons \cite{Scalapino1976}, a process quite
similar to the one we have studied. Our technique could also be useful to
describe quasiparticle damping in nuclear or neutron
matter~\cite{Ramos2001,Schuck2006}.

\begin{acknowledgments}
    This research was supported by the Bijzonder Onderzoeksfunds (BOF) of the
    University of Antwerp, the Fonds Wettenschappelijk Onderzoek Vlaanderen,
    project G.0429.15.N, and the European Union's Horizon 2020 research and
    innovation program under the Marie Sk\l odowska-Curie grant agreement number
    665501.
\end{acknowledgments}

\end{document}